\documentclass[reprint, aps, prl,superscriptaddress,floatfix]{revtex4-1}

\usepackage{amsmath,amssymb,mathtools}
\usepackage{bm}
\usepackage{graphicx}
\usepackage{dcolumn}
\usepackage{physics}
\usepackage{bbold}
\usepackage{xcolor}
\usepackage{comment}
\usepackage[hidelinks]{hyperref}

\newcommand{\PRLSec}[1]{\textit{#1}.}

\begin{document}

\title{Scaling of pseudospectra in exponentially sensitive lattices}

\author{
Ioannis Kiorpelidis and Konstantinos G. Makris \\
\textit{Department of Physics, University of Crete, 70013 Heraklion, Greece \\
and Institute of Electronic Structure and Laser (IESL), FORTH, 71110 Heraklion, Greece}
}

\begin{abstract}
One of the important features of non-Hermitian Hamiltonians is the existence of a unique type of singularities,
the so-called exceptional points (EPs). When the corresponding systems operate around such singularities,
they exhibit ultrasensitive behavior that has no analog in conservative systems. An alternative way to realize
such ultra-sensitivity relies on asymmetric couplings. Here we provide a comprehensive analysis based on
pseudospectra, that shows the origin of exponential sensitivity, without relying on topological zero modes or
the localization of all eigenstates (skin effect), but on the underlying extreme lattice non-normality. In particular,
we consider four different types of lattices (Hatano-Nelson, Sylvester-Kac, non-Hermitian Su-Schrieffer-Heeger
and a non-Hermitian random lattice) and identify the conditions for exponential sensitivity as a function of the
lattice size. Complex and structured pseudospectra reveal the signatures of exponential sensitivity both on the
eigenvalue spectra and on the underlying dynamics. Our study, may open new directions on studies related to the
exploitation of non-normality for constructing ultra-sensitive systems that do not rely on the existence of EPs.
\end{abstract}

\maketitle

%#################################################################################################################

\PRLSec{Introduction}
One of the cornerstones of non-Hermitian physics \cite{Moiseyev2011}, is the existence of the so-called exceptional points (EPs) \cite{Heiss2004,Berry2004}. Unlike conservative problems, where symmetries cause only eigenvalue degeneracy, here both eigenvalues and eigenmodes coalesce at specific control parameter values. The main physical consequence of eigenstate coalescence is the fact that weak perturbations can dramatically impact the system's response, which depends nonlinearly (algebraically) on the perturbation strength, in contrast to the Hermitian case, where the dependence is linear. This characteristic property of EPs has been studied in various contexts from mathematical physics to microwave cavities \cite{Heiss2004b,Richter2001}.
But their true potential has been reached in the framework of parity-time symmetric \cite{Bender1998} and non-Hermitian photonics \cite{PT0,PT1,PT2,PT3,PT4,PT5}, where EPs have been experimentally realized and led to various applications, such as ultrasensitive microresonators \cite{Wiersig2016,Wiersig2014,Khajavikhan2017,Yang2017}, exceptional point gyroscopes \cite{Khajavikhan2019}, and single mode nanolasers \cite{Khajavikhan2014}, to name a few. 

Another intriguing effect, characteristic of non-Hermitian Hamiltonians, is the non-Hermitian skin effect (NHSE)~\cite{Lee,review_NHSE,Torres2018}, which is a direct outcome of non-symmetric coupling between adjacent elements. This asymmetry is another alternative way to physically implement non-Hermiticity, with the other being the spatial combination of gain-loss materials. Counterintuitively, all the eigenmodes of such periodic systems are localized on one lattice end for Dirichlet boundary conditions. Such localization stems from the fact that any periodic lattice can be mapped to a Hatano-Nelson array under an imaginary gauge transformation \cite{Hatano1,Hatano2}. NHSE has been extensively studied not only in the framework of Hatano-Nelson Hamiltonians, but also in other lattice systems that exhibit asymmetric couplings. The experimental realization of Hatano-Nelson lattices, was elusive for long time due to the difficulty of realizing asymmetric couplings. Nevertheless, the NHSE was recently experimentally demonstrated in various photonic systems, like microring optical cavities, coupled optical fiber loop networks and mode locked lasers~\cite{Szameit2020,Khajavikhan2022,Gao2023,Nori2024}.

The latter studies regarding the NHSE (off-diagonal asymmetry), along with the initial ones regarding the spatial combination of gain-loss materials (on-diagonal complex elements), have established the framework of non-Hermitian photonics, where non-Hermiticity can be realized based on these two different methodologies. The synergy of non-Hermiticity with other physical effects, such as topological bands or optical nonlinearity, has widened the context of non-Hermitian photonics on both theoretical~\cite{Ueda2018,Sato2019} and experimental fronts~\cite{Szameit2017,Bandres2018,Chen2021}.

Similar to EPs, the NHSE provides an alternative way to realize ultrasensitive devices based on asymmetric couplings~\cite{Sato2020,Bergholtz2020,McDonald2020,Budich2022,Dong2022,zhang2024true,Sato2024}. In this regard, the exponential sensitivity associated with NHSE is attributed to its underlying topological nature~\cite{Bergholtz2020}, and has been recently experimentally demonstrated \cite{Zhang2023,Parto2023,Nori2024b,Achilleos2024}. In fact, the eigenspectrum of the problem changes dramatically depending on the boundary conditions, namely closed or periodic. So far all relevant studies devoted to topological sensors and topological amplification rely on perturbation theory and are strictly focused on the sensitivity of the zero eigenmode, by examining various mathematical sensitivity metrics such as energy shifts and condition numbers \cite{Sato2024}, for example. 

\begin{figure*}
\begin{center}
\includegraphics[width=2\columnwidth]{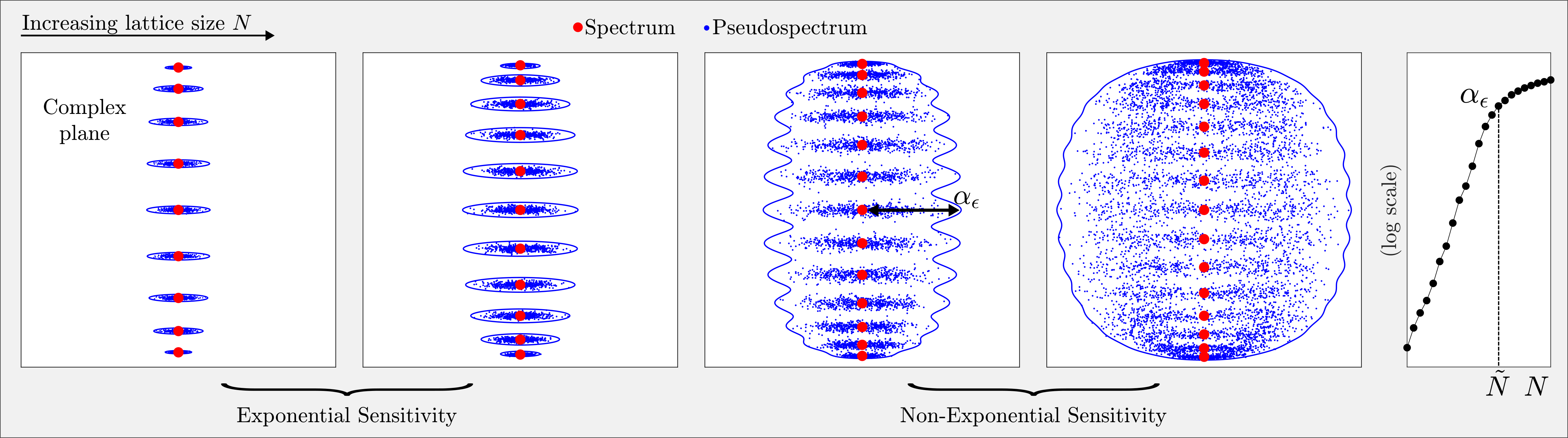}
\caption
{
Schematic of complex pseudospectrum scaling with the lattice size $N$. For small $N$’s, the pseudospectrum (blue dots) consists of
$N$ disks-clouds centered around the eigenvalues (red dots). By increasing $N$, the pseudospectrum extends in the complex plane and beyond a critical value $\tilde{N}$ it forms a single cloud. The abscissa $\alpha_{\epsilon}$ measures the extend of the pseudospectrum: it grows exponentially with $N$ up to $\tilde{N}$.
}
\label{fig0}
\end{center}
\end{figure*}

In this paper, we investigate the effect of exponential sensitivity beyond its topological context and the NHSE. In particular, we apply the theory of pseudospectra~\cite{Trefethen2005}, which, outside the fields of applied mathematics, fluid mechanics~\cite{Trefethen1993,Schmid2007}, and a few studies on excess noise in laser cavities~\cite{Longhi2000,Oppo2008}, remains largely unexplored in physics.
Despite this, pseudospectra have been introduced in the context of non-Hermitian optics, as they unify the concepts of spectral sensitivity and power dynamics \cite{Makris2014,Makris2024,Makris2021,Makris2022}. Furthermore, they offer a mathematically rigorous foundation for describing a range of problems involving topology and driven systems~\cite{Sato2023,Sato2020b,Kiorpelidis2024}, making them particularly relevant in modern non-Hermitian physics. 
Our results reveal two different aspects: From the mathematical point of view, the application of pseudospectra theory in this context allows us to move beyond perturbation theory and unify the spectral and dynamical sensitivity. 
From the physical point of view, which is our focus, the pseudospectra theory reveals the critical values of lattice size, of perturbation strength and of device length, that determine whether we have exponential sensitivity or not.
We consider different tight-binding lattices, with asymmetric nearest-neighbor hopping terms, like the Hatano-Nelson (HN)~\cite{Hatano1}, Sylvester-Kac (SK)~\cite{Todd1991,Willms2014}, non-Hermitian Su-Schrieffer-Heeger (SSH) model~\cite{Lieu2018,Wang2018}, and a disordered lattice with random coupling coefficients.
We study the scalings of complex~\cite{Trefethen2005} and structured pseudospectra \cite{Lubich2024_main, Sokolov2003,Graillat2006} with the lattice size $N$, and find that the pseudospectral abscissa $\alpha_{\epsilon}$ and the maximum power growth exhibit exponential sensitivity to $N$ in all cases, implying that tuning this parameter is crucial for practical sensing applications.
Figure~\ref{fig0} illustrates how the pseudospectrum depends on $N$ and also shows the determination of $\alpha_{\epsilon}$ along with its variation with $N$.

%#################################################################################################################

\begin{figure}
\begin{center}
\includegraphics[width=1\columnwidth]{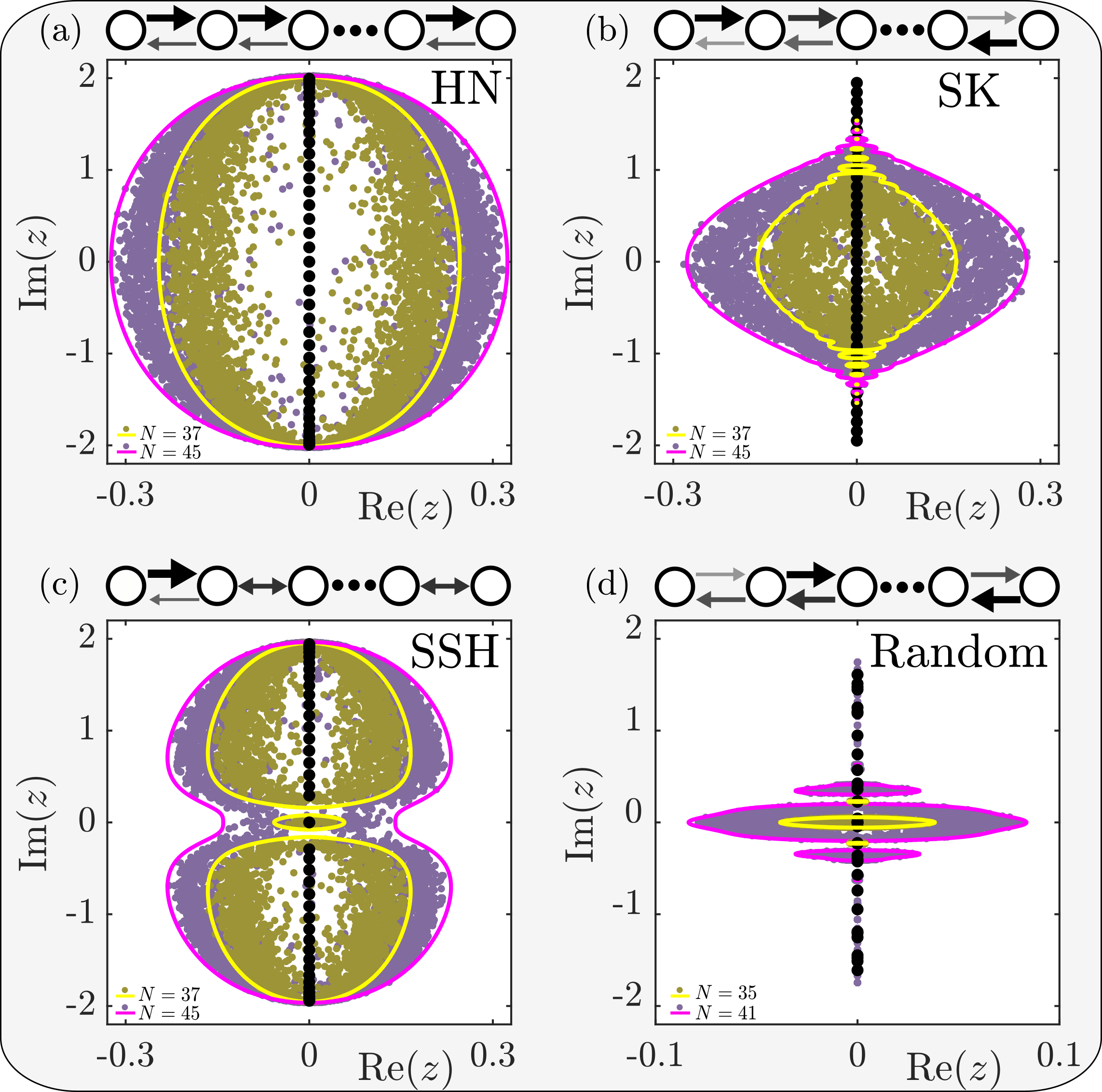}
\caption
{
(a)-(d)
Eigenspectra on the complex plane $z \in \mathbb{C}$ (black
dots) and complex pseudospectra of the matrix $\mathbf{A}= -i\mathbf{H}$, for $ \epsilon= 10^{-5}$ and for two lattice sizes $N$ (purple and yellow dots/lines).
(a) Hatano-Nelson model ($J_R = 3/2$, $J_L = 2/3$),
(b) Sylvester-Kac model ($c = 2$), 
(c) Non-Hermitian Su-Schrieffer-Heeger model
($J_R = 1.8$, $J_L = 0.4$, $J= 1.1$) and 
(d) Random asymmetric model (hopping terms $J_n$ and
$\tilde{J}_n$ uniformly in $[0,2]$).
}
\label{fig1}
\end{center}
\end{figure}

\PRLSec{Non-Hermitian lattices}
Let us start by presenting in Figs.~\ref{fig1}(a)-(d) the four discrete lattices that we consider in our study. 
The wave evolution is governed by the coupled mode equation of paraxial optics  for the electric field's envelope amplitudes-${\boldsymbol{\psi}}$, namely
$\dot{\boldsymbol{\psi}}=-i \mathbf{H} \boldsymbol{\psi}$, with $\boldsymbol{\psi}(t=0)=\boldsymbol{\psi}_0$,
where dot denotes differentiation with respect to time $t$ (for cavities) or propagation distance (for waveguides), $\mathbf{H}$ the Hamiltonians of the models, and $\boldsymbol{\psi}_0$ the initial conditions.
The Hamiltonians can all be written in the following form: 
$
\mathbf{H} = -\sum_{n=1}^{N-1} \left( J_n |n\rangle \langle n+1| + \tilde{J}_n |n+1\rangle\langle n| \right),
$
where $|n\rangle$ is the basis state localized at lattice site $n$, $J_n$ is the hopping amplitude for transitions from site $n+1$ to site $n$ ($n=1,...,N$), and $\tilde{J}_n$ is the hopping amplitude in the reverse direction.
In the Hatano-Nelson model $J_n=J_L$ and $\tilde{J}_n=J_R$ $\forall n$, 
in the Sylvester-Kac $J_n=cn/N$ and $\tilde{J}_n=c(N-n)/N$,
in the SSH $J_n=t_L$ and $\tilde{J}_n=J_R$ for odd $n$ and $J_n=\tilde{J}_n=t$ for even $n$, and in the random model $J_n$ and $\tilde{J}_n$ are random numbers uniformly distributed in $[0,h]$.

Due to the asymmetry of the hopping terms, $J_n \neq \tilde{J}_n$, the Hamiltonians $\mathbf{H}$ exhibit different right and left eigenvectors, $\boldsymbol{\psi}_n^R$ and $\boldsymbol{\psi}_n^L$ respectively, $n=1,2,...,N$, with distinct spatial profiles; see Supplemental Material (SM).
We note here that in all models, we can find at least one pair of right/left eigenstates which are localized at opposite lattice edges (for the random model we have selected a realization that satisfies this property) and thus their spatial overlap depends exponentially on the lattice size.

\begin{figure*}
\begin{center}
\includegraphics[width=2\columnwidth]{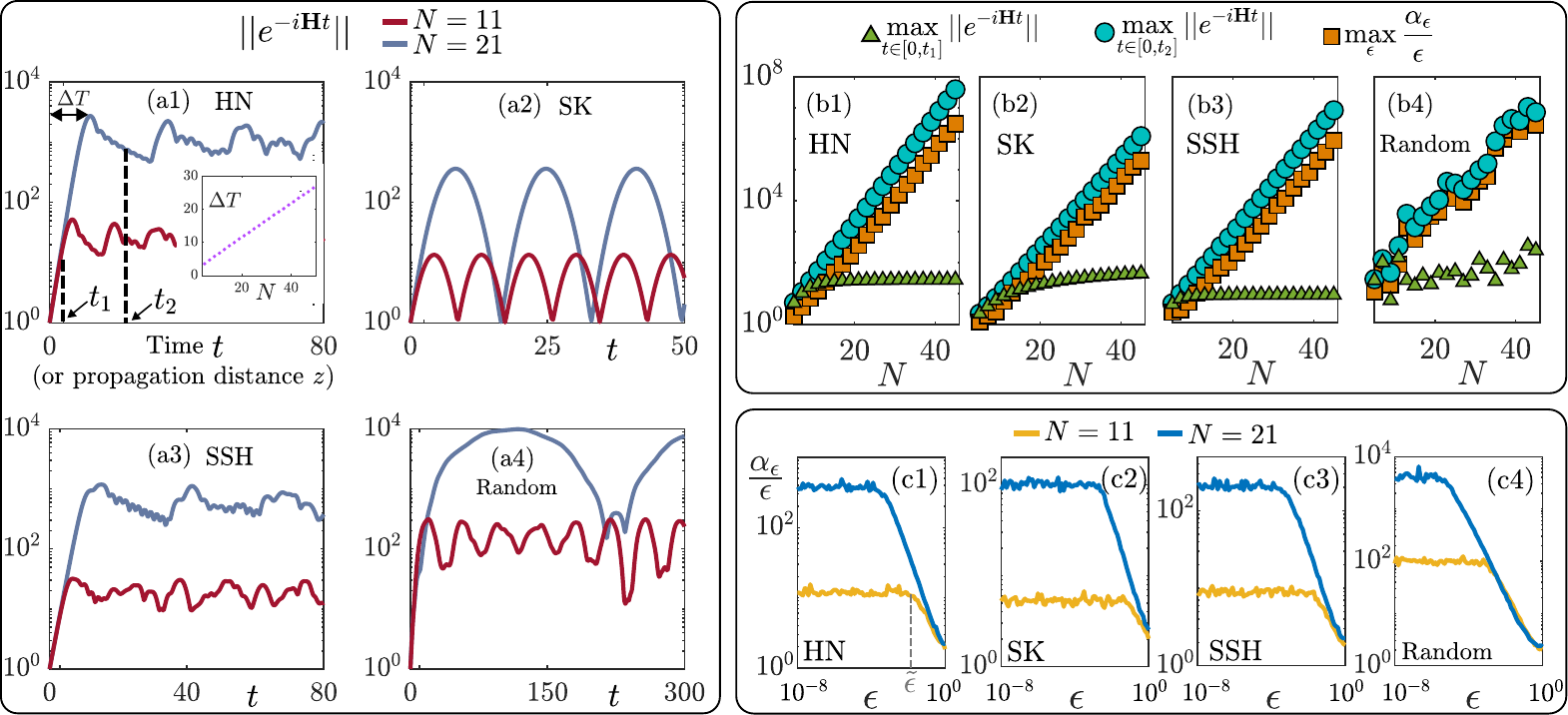}
\caption
{
Scaling of spectral and dynamical sensitivity based on complex pseudospectra. (a1)–(a4) Time-evolution of the propagator norm for two lattice sizes ($N=11$ and $N=21$).
Inset in panel (a1): The time $\Delta T$ scales linearly with $N$.
(b1)–(b4) Shown are the $\max_{t \in [0,t_1]} ||e^{-i\mathbf{H}t}||$ for $t_1 \ll 1$, the $\max_{t \in [0,t_2]} ||e^{-i\mathbf{H}t}||$ for $t_2 \gg 1$, and the Kreiss constant $\max_{\epsilon} (\alpha_{\epsilon}/\epsilon)$.
 The first quantity does not scale exponentially with $N$, whereas
the latter two do.
In  (b1) $t_1=4$, in (b2) $t_1=2.5$, in (b3) $t_1=3$ and in (b4) $t_1=10$. In all panels $t_2=800$.
(c1)–(c4)
Variation of $\alpha_{\epsilon}/\epsilon$ with $\epsilon$, for two different lattice sizes ($N=11$ and $N=21$).
Notice that the abscissa grows linearly up to a value $\tilde{\epsilon}$, while beyond $\tilde{\epsilon}$ it scales as $\alpha_{\epsilon} \sim \epsilon^{\delta}$.
The value of the abscissa at $\tilde{\epsilon}$ is an estimate for $\Delta T$.
}
\label{fig2}
\end{center}
\end{figure*}

The solution of the coupled mode equation is given by $\boldsymbol{\psi}(t)=e^{-i\mathbf{H}t} \boldsymbol{\psi}_0$,
where the matrix $e^{-i\mathbf{H}t}$ is refereed as the propagator or evolution matrix \cite{Trefethen2005}.
The quantity  $||\boldsymbol{\psi}(t)||^2$ is a measurable observable of the system called the optical power and therefore the ratio $||\boldsymbol{\psi}(t)||/||\boldsymbol{\psi}_0||$ is  the square root of the amplification ratio of the lattice \cite{Makris2021, Makris2024}.
In all the considered lattices,  the eigenvalue spectra $\lbrace E_n \rbrace$, denoted  by $\sigma(\mathbf{H})$, are purely real regardless of whether $N$ is even or odd, since $\mathbf{H}$ can be mapped to a hermitian Hamiltonian by applying a similarity transformation (see SM).
The solution $\boldsymbol{\psi}(t)$ written in the eigenbasis $\lbrace  \boldsymbol{\psi}_n^R \rbrace$ of $\mathbf{H}$ takes the form, $\boldsymbol{\psi}(t)=\sum_{n=1}^N c_n e^{-i E_n t} \boldsymbol{\psi}^R_n$,
where the $c_n$ coefficients can be uniquely determined by the initial conditions $\boldsymbol{\psi}_0$. The eigenvalues $E_n$ are purely real (despite the non-Hermiticity of the matrix), and therefore the Euclidean norms $||\boldsymbol{\psi}(t)||$ are bounded in time.
In fact, the Hamiltonians $\mathbf{H}$ of the considered models are non-normal, i.e. $[\mathbf{H},\mathbf{H}^{\dagger}] \neq 0$,
and due to this any initial condition can in principle be amplified, in the sense that $||\boldsymbol{\psi}(t)||$ can exceed $||\boldsymbol{\psi}_0||$. In other words, the optical power $\sum_n|\psi_n|^2$ is not a conserved quantity  anymore but oscillates \cite{Makris2024,Makris2021}.

%#################################################################################################################

\PRLSec{Complex pseudospectra}
In order to systematically explore these power oscillations, we will apply the pseudospectra theory \cite{Trefethen2005}.
In this paragraph, we provide the simplest definition of complex pseudospectrum. Given any $\epsilon > 0$, the pseudospectrum $\sigma_{\epsilon}(\mathbf{A})$ of a matrix $\mathbf{A}$ is the set of complex numbers $z$ that are eigenvalues of perturbed matrices $\mathbf{A}+\mathbf{E}$, with $\mathbf{E}$ being a complex random matrix, whose 2-norm is less than $\epsilon$ \cite{comment_norm}. 
Strictly speaking the definition is,
$
\sigma_{\epsilon}(\mathbf{A})
=
\bigcup_{||\mathbf{E}|| < \epsilon}
\sigma(\mathbf{A}+\mathbf{E})
.
$
We present our results regarding the complex pseudospectra of the four lattices in Fig.~\ref{fig1}. More specifically, in Figs.~\ref{fig1}(a)-\ref{fig1}(d) we show the spectra (black dots) of the matrix $\mathbf{A}=-i\mathbf{H}$, as well as, the corresponding pseudospectra by computing 100 perturbed matrices $\mathbf{A}+\mathbf{E}$ for $\epsilon=10^{-5}$, and for two different lattice sizes $N$ (yellow and purple dots respectively).
Notice that the pseudospectrum $\sigma_{\epsilon}(\mathbf{A})$ forms a cloud of eigenvalues around the spectrum $\sigma(\mathbf{A})$ and its boundary [solid lines shown in Figs.~\ref{fig1}(a)-(d)] is determined by an equivalent pseudospectrum definition based on the resolvent of $\mathbf{A}$ (see SM).

%#################################################################################################################

\PRLSec{Scaling of complex pseudospectra  with $N$}
Regarding the spectral sensitivity, Fig.~\ref{fig0} presents a schematic depiction of the general behavior of all the four lattices that we consider. In particular, for small $N$'s the pseudospectrum consists of $N$ disks-clouds centered at the $N$ eigenvalues of $\mathbf{A}$. As $N$ increases, the radii of these disks grow, and beyond a certain value $\tilde{N}$, they merge into a single cloud.
A metric for the extent of the pseudospectrum is provided by the pseudospectral abscissa, $\alpha_{\epsilon}=\max \Re [\sigma_{\epsilon}(\mathbf{A})]$, which also provides a measure of the lattice’s spectral sensitivity. The right panel of Fig.~\ref{fig0} shows the variation of $\alpha_{\epsilon}$ with $N$: notice that it increases exponentially up to $\tilde{N}$ corresponding to the onset of the single-cloud regime.

Regarding the dynamical sensitivity of the lattice, a measure quantifying the power dynamics is the maximum growth at any given time $t$, $\max_{\boldsymbol{\psi}_0, \boldsymbol{||\psi}_0||=1} {||\boldsymbol{\psi}(t)||}{/||\boldsymbol{\psi}_0||}$. By definition, the latter quantity is equal to the 2-norm of the propagator matrix, which in turn is equal to its maximum singular value. Shown in Figs.~\ref{fig2}(a1)-(a4) is the time-evolution of the propagator norm for all models.
Despite the real spectrum of $\mathbf{H}$, amplification occurs ($||e^{-i\mathbf{H}t}||>1$) because $\mathbf{H}$ is non-normal.
Interestingly, for large times $t_2 \gg 1$, the maximum power growth scales exponentially with the lattice size $N$.
Let us note here that this exponential dependence is not constrained by a critical value of $N$, highlighting the distinction between exponential spectral sensitivity---bounded by a critical value $\tilde{N}$---and dynamical exponential sensitivity.
This is illustrated in Figs.~\ref{fig2}(b1)-(b4) where the $ \max_{t \in [0,t_2]} ||e^{-i\mathbf{H}t}||$ is displayed.
On the contrary, for small times $t_1 \ll 1$, the $ \max_{t \in [0,t_1]} ||e^{-i\mathbf{H}t}||$ does not scale exponentially with $N$ [see Figs.~\ref{fig2}(b1)-(b4)].
The corresponding time scale $\Delta T$ after which the lattice displays exponential sensitivity, scales linearly with $N$ [see the inset of Fig.~\ref{fig2}(a1)].
The pseudospectrum renders back information regarding the exponential sensitivity of the power dynamics through the Kreiss matrix theorem \cite{Trefethen2005}
$\mathcal{K}(\mathbf{A}) \leq    \max_t ||e^{\mathbf{A}t}|| \leq eN\mathcal{K}(\mathbf{A})$,
where $e$ is the Euler's number and $\mathcal{K}(\mathbf{A})=\max_{\epsilon >0} (\alpha_{\epsilon} / {\epsilon})$ is the Kreiss constant.
Figures~\ref{fig2}(b1)-(b4) show that $\mathcal{K}(\mathbf{A})$ increases exponentially with $N$.

\PRLSec{Scaling of complex pseudospectra with $\epsilon$}
In this paragraph, we are interested to see the scaling of the pseudospectral abscissa $\alpha_{\epsilon}$ of  $\mathbf{A}$, by keeping $N$ constant and varying the perturbation parameter $\epsilon$.
We present our results in Figs.~\ref{fig2}(c1)–(c4).
Notice that $\alpha_{\epsilon}$ increases linearly up to a value $\tilde{\epsilon}$, and for higher values it scales as $\alpha_{\epsilon} \sim \epsilon^{\delta}$.
In SM we show that the exponent $\delta$ converges to $1/N$ for large asymmetry between right and left couplings, suggesting that the lattice exhibits the same sensitivity with that of an exceptional point of order $N$.
This scaling of the pseudospectral abscissa is independent of the lattice size $N$,
yet the value $\tilde{\epsilon}$ decreases with increasing $N$. 
Let us make a comment at this point:
The pseudospectrum expands in the complex plane, not only by increasing the lattice size $N$  as is shown in Figs.~\ref{fig0} and \ref{fig1}, but also by increasing the perturbation parameter $\epsilon$. As the perturbation parameter $\epsilon$ increases , the pseudospectrum expands in the complex plane as well, and a similar transition from $N$ disks-clouds to a single-cloud occurs. This transition
takes place at the critical value $\tilde{\epsilon}$. Moreover, the value of $1/\tilde{\alpha}_{\epsilon}$ at $\tilde{\epsilon}$ offers estimate for the time scale $\Delta T$  \cite{Trefethen2005} after which the lattice supports exponential sensitivity with $N$.

\begin{figure}
\begin{center}
\includegraphics[width=1\columnwidth]{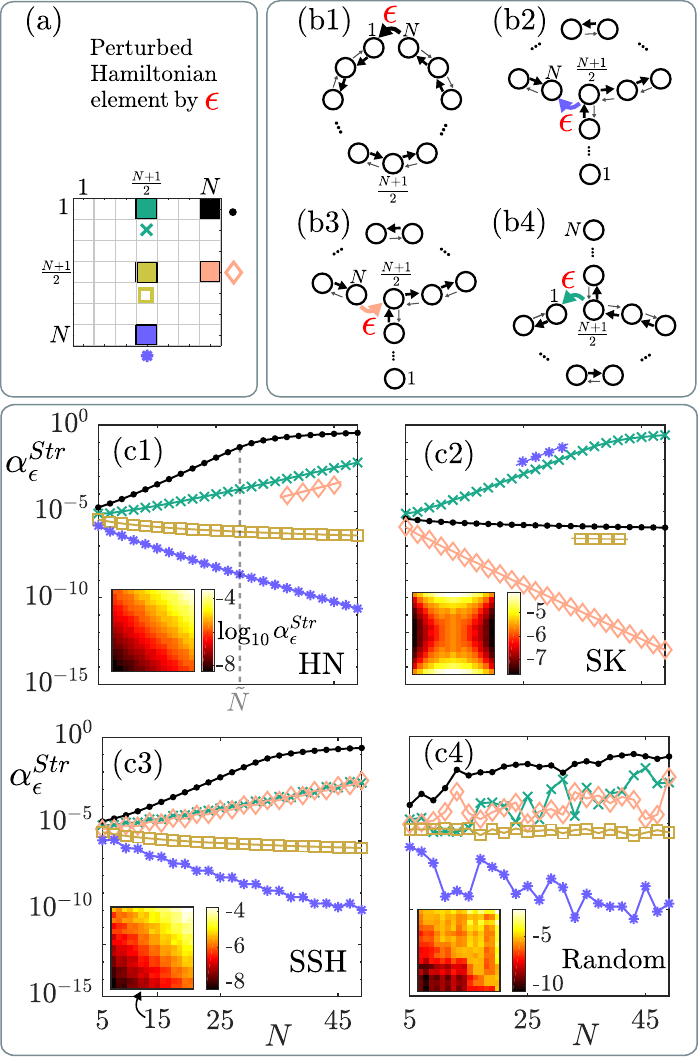}
\caption
{Scaling of spectral sensitivity with $N$, based on the structured pseudospectra. (a) The matrix denotes five different elements of $\mathbf{H}$ that are perturbed by a complex random number of $\epsilon$ magnitude.
(b1)-(b4) Different geometrical configurations resulting from the perturbation of different Hamiltonian elements denoted in (a).
(c1)-(c4) Variation of the structured pseudospectral abscissa $\alpha_{\epsilon}^{\text{Str}}$ with $N$,
when each one of the five different elements of $\mathbf{H}$ denoted in (a) is perturbed. $\epsilon$ is set at $10^{-5}$.
The insets matrices illustrate which Hamiltonian element demonstrates the highest sensitivity when perturbed.}
\label{fig3n}
\end{center}
\end{figure}

%#################################################################################################################

\PRLSec{Scaling of structured pseudospectra with $N$}
As stated before, the abscissa $\alpha_{\epsilon}$ of the complex pseudospectrum scales exponentially with lattice size $N$, up to a critical value $\tilde{N}$.
Such a scaling, however, involves a perturbation matrix $\mathbf{E}$ of which all elements are generally nonzero
-- it affects all entries of the Hamiltonian --
making it unsuitable for studying the sensitivity of the lattices's spectra under experimentally realistic perturbations.
The latter analysis can instead be performed with the structured pseudospectrum \cite{Lubich2024_main,Sokolov2003, Graillat2006}:
the structured pseudospectrum of a matrix $\mathbf{A}$, denoted as $\sigma_{\epsilon}^{\text{Str}}(\mathbf{A})$, is defined analogously to the complex pseudospectrum with the difference that the random matrix $\mathbf{E}$ exhibits specific structure,
$
\sigma_{\epsilon}^{\text{Str}}(\mathbf{A})
=
\bigcup_{\mathbf{E}-\text{Structured},||\mathbf{E}|| < \epsilon}
\sigma(\mathbf{A}+\mathbf{E})
.
$
For instance, $\mathbf{E}$ may have nonzero elements exclusively on its diagonal, suggesting that only the onsite potential of the lattice is perturbed (see SM for more details and illustrations of the structured pseudospectrum).

When the Hamiltonian $\mathbf{H}$ is subject to a  single-element perturbation [Fig.~\ref{fig3n}(a)], it is suggested that two lattice sites get connected.
Figures \ref{fig3n}(b1)-(b4) illustrate how perturbing four distinct elements of $\mathbf{H}$  leads to the emergence of four unique geometric configurations.
Moreover, as can be seen in Figs.~\ref{fig3n}(c1)-(c4), the $\alpha_{\epsilon}^{\text{Str}}$ scales exponentially with the system size $N$ when perturbing appropriate Hamiltonian elements.
Based on the structured pseudospectra technique, we can identify which element results to the most sensitive response of the whole system.
For instance, for the inset plots of Figs.~\ref{fig3n}(c1)-(c4), we kept both $\epsilon$ and $N$ fixed ($\epsilon=10^{-5}$ and $N=15$), we considered single element perturbation matrices $\mathbf{E}=\epsilon \delta_{n,m}$ with $n,m=1,...,15$, and we computed the value of the structured abscissa (colorbar) for each pair $(n,m)$. At this point, we  should stress the difference betweeen the structured pseudospectrum approach and the conventional perturbation methods \cite{Bergholtz2020}.
In the latter ones, the focus is on the energy shift of one individual zero-energy mode under applied perturbations, whereas in the former one, the collective effect of all eigenvalues is taken into account.

%#################################################################################################################

\PRLSec{Discussion-Conclusions}
In summary, we have presented a scaling analysis of pseudospectra in exponential sensitive lattices. In particular, we considered four different examples of coupled optical elements (cavities or waveguides) that have asymmetric coupling coefficients. On one hand, the scaling of complex pseudospectra (of both the unperturbed and perturbed Hamiltonian) provides us useful information about the power dynamics scaling with the system's size $N$. Moreover, additional scaling with respect to the perturbation strength $\epsilon$, gives us an understanding of the limits of this exponential sensitivity. On the other hand, scaling of the structured pseudospectra offers valuable information regarding the  sensitivity of the spectrum, again with respect to $N$. 

Unlike all previous works, we do not rely on topological effects driven from the extreme sensitivity on boundary conditions, but we focus on characterizing the underlying non-normality. Let us note here the strengths of pseudospectra: (a) the computation of eigenvalue spectra could be numerically much more efficient, both in terms of computational cost and stability, than the computation of the matrix exponentials. Thus especially for problems involving large matrices (for example quantum problems) the pseudospectra could offer a significant advantage. (b) Perturbation theory focuses on the impact of a perturbation on individual eigenvalues, while the pseudospectrum reveals the collective impact of a perturbation on all eigenvalues; it provides a global perspective.  This global framework can be utilized by machine learning to design ultrasensitive structures.

In conclusion, while recent experiments in non-Hermitian photonics have demonstrated asymmetric optical lattices \cite{Szameit2020, Khajavikhan2022, Nori2024, Parto2023}, the realization of exponential sensitivity in power dynamics, as well as, its demonstration in non-topological lattices remains an open problem. Such a demonstration could, in principle, be achieved through discrete-time quantum walks of optical pulses in coupled fiber loops \cite{Szameit2020}. Thus, our results may pave the way for efficient design of ultrasensitive devices without relying on higher order exceptional points.

%#################################################################################################################

\PRLSec{Acknowledgments}
K.G.M. acknowledges enlightening discussions with Professsor C. Lubich during the workshop “Nonlinear Optics: Physics, Analysis, and Numerics” at Mathematisches Forschungsinstitut Oberwolfach (11–15 March 2024), regarding the structured pseudospectra. This  project was funded by the European Research Council 
(ERC-Consolidator) under Grant Agreement No. 101045135
(Beyond\_Anderson).

%#################################################################################################################

% ================== SUPPLEMENT START =====================
\clearpage                 % start on a new page
\onecolumngrid

\begin{center}
  {\large \textbf{Supplemental Material: Scaling of pseudospectra in exponentially sensitive lattices}}\par
  \vspace{8pt}
  Ioannis Kiorpelidis and Konstantinos G. Makris\par
  \vspace{2pt}
  \textit{Department of Physics, University of Crete, 70013 Heraklion, Greece\\
  and Institute of Electronic Structure and Laser (IESL), FORTH, 71110 Heraklion, Greece}\par
  \vspace{12pt}
\end{center}

%#####################################################################################################################################

\section{Hamiltonian matrix representations of our four models}
We give here the Hamiltonian matrices of the models used in this work [Hatano-Nelson (HN), Sylvester-Kac (SK), non-Hermitian Su-Schrieffer-Heeger (SSH) and a non-Hermitian random lattice]
\begin{equation}
\begin{aligned}
H_{\text{HN}} =
-
    \begin{pmatrix}
        0 & J_L & 0 & 0 & \cdots & 0 & 0 \\
        J_R & 0 & J_L & 0 & \cdots & 0 & 0 \\
        0 & J_R & 0 & J_L & \cdots & 0 & 0 \\
        \vdots & \vdots & \vdots & \vdots & \ddots & \vdots & \vdots \\
        0 & 0 & 0 & 0 & \cdots & 0 & J_L \\
        0 & 0 & 0 & 0 & \cdots & J_R & 0 \\
    \end{pmatrix}_{N \times N}
~~~
H_{\text{SK}} =
-
    c
    \begin{pmatrix}
        0 & \frac{1}{N} & 0 & 0 & \cdots & 0 & 0 \\
        \frac{N-1}{N} & 0 & \frac{2}{N} & 0 & \cdots & 0 & 0 \\
        0 & \frac{N-2}{N} & 0 & \frac{3}{N} & \cdots & 0 & 0 \\
        \vdots & \vdots & \vdots & \vdots & \ddots & \vdots & \vdots \\
        0 & 0 & 0 & 0 & \cdots & 0 & \frac{N-1}{N} \\
        0 & 0 & 0 & 0 & \cdots & \frac{1}{N} & 0 \\
    \end{pmatrix}_{N \times N}
    \\\\
     H_{\text{SSH}} =
     -
    \begin{pmatrix}
        0 & J_L & 0 & 0 & \cdots & 0 & 0 \\
        J_R & 0 & J & 0 & \cdots & 0 & 0 \\
        0 & J & 0 & J_L & \cdots & 0 & 0 \\
        \vdots & \vdots & \vdots & \vdots & \ddots & \vdots & \vdots \\
        0 & 0 & 0 & 0 & \cdots & 0 & J_L \\
        0 & 0 & 0 & 0 & \cdots & J_R & 0 \\
    \end{pmatrix}_{N \times N}
    ~~~
    H_{\text{Random}} =
    -
    \begin{pmatrix}
        0 & J_1 & 0 & 0 & \cdots & 0 & 0 \\
        \tilde{J}_1 & 0 & J_2 & 0 & \cdots & 0 & 0 \\
        0 & \tilde{J}_2 & 0 & J_3 & \cdots & 0 & 0 \\
        \vdots & \vdots & \vdots & \vdots & \ddots & \vdots & \vdots \\
        0 & 0 & 0 & 0 & \cdots & 0 & J_{N-1} \\
        0 & 0 & 0 & 0 & \cdots & \tilde{J}_{N-1} & 0 \\
    \end{pmatrix}_{N \times N}
    .
\end{aligned}
\label{matrices}
\end{equation}
$N$ is the lattice size and the elements on the subdiagonal and lower diagonals are the hopping amplitudes between neighbor sites of the lattice.
The entries on the main diagonals are zero because we have assumed identical elements.

We can prove that the eigenvalues of the matrices given in Eq.~\eqref{matrices} are purely real by employing a similarity transformation as follows:
First, we define the transformation matrix $\mathbf{D}=\text{diag}(\delta_1,\delta_2,...,\delta_N)$ with $\delta_1=1$ and $\delta_{n}=\sqrt{\frac{\tilde{J}_{n-1} \cdots \tilde{J}_{1}}{{J}_{n-1} \cdots {J}_{1}}}$ for $n=2,...,N$.
Then we show that the matrix $\mathbf{H}'=\mathbf{D}^{-1}\mathbf{H} \mathbf{D}$ is hermitian.
Specifically, it is a tridiagonal matrix with equal upper and lower diagonals and zero diagonal.
The entries in the upper and lower diagonals are equal to $\sqrt{\tilde{J}_n J_n}$ with $n=1,...,N-1$.
Since $\mathbf{H}'$ is hermitian, it possesses real eigenvalues, and since $\mathbf{H}$ and $\mathbf{H}'$ are similar, they have the same eigenvalues.
Moreover, the matrices given in Eq.~\eqref{matrices}, possess a zero eigenvalue for odd $N$.

\begin{figure}[h!]
\begin{center}
\includegraphics[width=1\columnwidth]{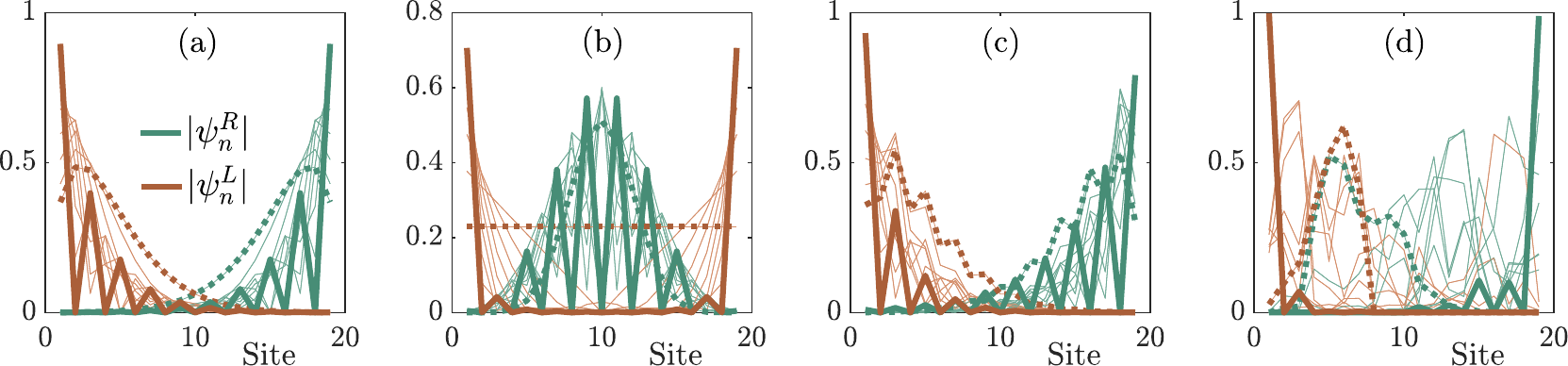}
\caption
{
Eigenstates of the four different models. Amplitudes of the right (green), left (brown) eigenstates of the four lattices (for $N=19$):
(a) Hatano-Nelson model ($J_R=3/2$, $J_L=2/3$),
(b) Sylvester-Kac model ($c=2$),
(c) non-Hermitian Su-Schrieffer-Heeger model ($J_R=1.8$, $J_L=0.4$, $J=1.1$),
(d) random model (hopping terms $J_n$ and $\tilde{J}_n$ uniformly in [0,2]).
In all cases, bold (dashed) lines indicate the zero-energy (lowest-energy) modes.
}
\label{supple_1}
\end{center}
\end{figure}

Due to the asymmetry of the hopping terms, the Hamiltonians $\mathbf{H}$ exhibit different right and left eigenvectors, $\boldsymbol{\psi}_n^R$ and $\boldsymbol{\psi}_n^L$ respectively, $n=1,2,...,N$, with distinct spatial profiles (see Fig.~\ref{supple_1}).
In particular, in the HN and SSH models all the right/left eigenvectors are localized at opposite edges of the lattice, whereas in the SK model two eigenvectors are extended, and in the random model the eigenvectors are localized anywhere in the lattice; in the latter model, for each $N$, we selected a realization where the right (left) eigenstate corresponding to zero energy is localized at the right (left) lattice end.

%#####################################################################################################################################

\section{Pseudospetrum definition based on the resolvent}

In addition to the definition of the complex pseudospectrum based on perturbed matrices that is mentioned in the main text,  there is an equivalent definition based on the resolvent of the matrix $\mathbf{A}$ \cite{Trefethen2005_supple}:
the complex pseudospectrum is the set of complex numbers $z$ satisfying
\begin{equation}
    \sigma_{\epsilon}(\mathbf{A})= \lbrace z \in \mathcal{C}: ||(z\mathbf{I}-\mathbf{A})^{-1}||>\epsilon^{-1} \rbrace
\end{equation}
where $\mathbf{I}$ is the identity matrix and $\mathbf{R}(z)=(z\mathbf{I}-\mathbf{A})^{-1}$ is the resolvent of $\mathbf{A}$.

%#####################################################################################################################################

\section{Exponential localization of right/left eigenstates}

We provide here arguments showing that a necessary condition for a lattice to exhibit exponential sensitivity with the size $N$, is that at least one pair of right and left eigenstates of the Hamiltonian are exponentially localized at two different lattice sites, and the distance between the two corresponding localization centers is $O(N)$.

\begin{figure}[h!]
\begin{center}
\includegraphics[width=0.95\columnwidth]{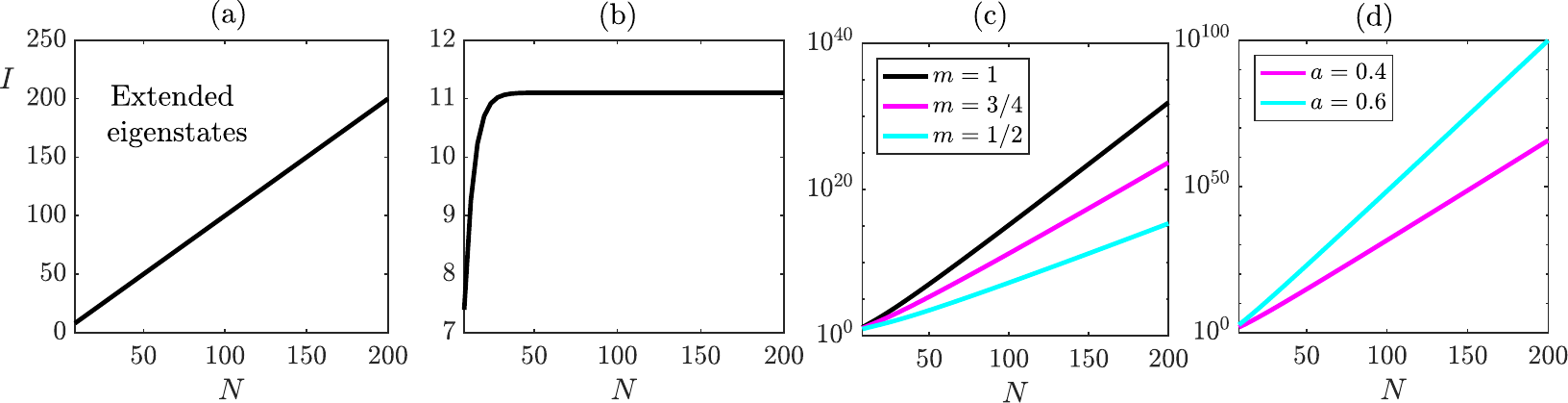}
\caption
{
The quantity 
$I=\boldsymbol{\psi}^{\dagger}_0 
    \boldsymbol{w}_i
    \boldsymbol{u}_i^{\dagger}
    \boldsymbol{u}_i
    \boldsymbol{w}_i^{\dagger}
    \boldsymbol{\psi}_0
$
as a function of lattice size $N$,
for 
$\boldsymbol{u}$ and $\boldsymbol{w}$ given in Eq.~\eqref{appe4}, and initial state
$ 
\boldsymbol{\psi}_0
=
\begin{pmatrix}
1&\cdots&1
\end{pmatrix}^{\dagger}
$
(a) $a=0$.
(b) $a=0.2$, $n_2=1$ and $n_1=3$
(c) $a=0.2$, $n_2=1$ and $n_1=mN$ with $m=1$ (black curve), $m=3/4$ (magenta curve), $m=1/2$ (cyan curve) and $m=1/4$ (green curve).
(d) $n_2=1$, $n_1=N$ and $a=0.4$ (magenta curve), $a=0.6$ (cyan curve).
}
\label{fig10_2}
\end{center}
\end{figure}

The time-evolution of an initial state $\boldsymbol{\psi}_0$ is governed by
\begin{equation}
    \dot{\boldsymbol{\psi}}=\mathbf{A} \boldsymbol{\psi}
    \label{appe1}
\end{equation}
where $\mathbf{A}=-i\mathbf{H}$ and $\mathbf{H}$ is the Hamiltonian. At time $t=0$, $\boldsymbol{\psi}(t=0)=\boldsymbol{\psi}_0$.
The solution of Eq.~\eqref{appe1} is given by $\boldsymbol{\psi}(t)=e^{\mathbf{A}t}\boldsymbol{\psi}_0$ and therefore the norm of the vector $\boldsymbol{\psi}$ evolves according to
\begin{equation}
    ||\boldsymbol{\psi}(t)||= \boldsymbol{\psi}^{\dagger}_0e^{\mathbf{A}^{\dagger}t}e^{\mathbf{A}t} \boldsymbol{\psi}_0.
    \label{appe2}
\end{equation}
The matrix $\mathbf{A}$ is non-normal and therefore it possesses distinct right and left eigenvectors, $\boldsymbol{u}_n$ and $\boldsymbol{w}_n$ respectively, with $n=1,...,N$.
The $\boldsymbol{u}_n$'s $\boldsymbol{w}_n$'s are non-orthogonal among themselves due to the non-normality of $\mathbf{A}$, yet they satisfy the orthogonality condition $\boldsymbol{w}_m^{\dagger} \boldsymbol{u}_n=\delta_{nm}$.
Using the spectral decomposition of $\mathbf{A}=\sum_{n=1}^{N}\lambda_n \boldsymbol{u}_n \boldsymbol{w}_n^{\dagger}$, where $\lambda_n$ is the eigenvalue corresponding to $\boldsymbol{u}_n$ and $\boldsymbol{w}_n$ (assuming no degeneracies), then Eq.~\eqref{appe2} is written in the following form
\begin{equation}
    \displaystyle
    ||\boldsymbol{\psi}(t)||= 
    \sum_{n=1}^N
    \sum_{m=1}^N
    \boldsymbol{\psi}^{\dagger}_0 
    \boldsymbol{w}_n
    \boldsymbol{u}_n^{\dagger}
    \boldsymbol{u}_m
    \boldsymbol{w}_m^{\dagger}
    \boldsymbol{\psi}_0
    e^{t(\lambda_n+\lambda_m^*)}
    .
    \label{appe4}
\end{equation}
Let us focus on one term of the latter sum, of the form 
$I=\boldsymbol{\psi}^{\dagger}_0 
    \boldsymbol{w}_i
    \boldsymbol{u}_i^{\dagger}
    \boldsymbol{u}_i
    \boldsymbol{w}_i^{\dagger}
    \boldsymbol{\psi}_0$
. For convenience we will drop the subscript $i$ in the following.
We assume that $\boldsymbol{u}$ and $\boldsymbol{w}$ are exponentially localized around the sites $n_1$ and $n_2$ respectively, having the form
\begin{equation}
    \boldsymbol{u}
    =
    M
    \begin{pmatrix}
        e^{-a|1-n_1|}\\
        e^{-a|2-n_1|}\\
        \vdots \\
        e^{-a|N-1-n_1|}\\
        e^{-a|N-n_1|}
    \end{pmatrix}
    ~~~\text{and}~~~
    \boldsymbol{w}
    =
    \begin{pmatrix}
        e^{-b|1-n_2|}\\
        e^{-b|2-n_2|}\\
        \vdots \\
        e^{-b|N-1-n_2|}\\
        e^{-b|N-n_2|}
    \end{pmatrix}
    \label{appe4}
\end{equation}
where $a$ and $b$ are localization scales; without loss of generality we set $a=b$. The factor $M$ is given by the normalization condition 
$\boldsymbol{w}^{\dagger} \boldsymbol{u}=1$.
In Fig.~\ref{fig10_2} we show the quantity 
$I=\boldsymbol{\psi}^{\dagger}_0 
    \boldsymbol{w}_i
    \boldsymbol{u}_i^{\dagger}
    \boldsymbol{u}_i
    \boldsymbol{w}_i^{\dagger}
    \boldsymbol{\psi}_0$, as a function of $N$, for the following cases (without loss of generality we chose as an initial state the 
$ 
\boldsymbol{\psi}_0
=
\begin{pmatrix}
1&\cdots&1
\end{pmatrix}^{\dagger}
$
):
In panel (a) we chose $a=0$, meaning that the eigenstates are extended.
Notice that $I$ scales linearly with $N$.
In panel (b) we set $a=0.2$, $n_2=1$ and $n_1=3$, meaning that the two eigenstates are localized, yet their localization centers are fixed at two specific lattice sites and therefore the distance between these centers is $d_{n_1,n_2}=c$ for all $N$ (for the chosen parameters $c=2$).
Notice that $I$ does not increase exponentially with increasing $N$ in this case as well.
In panel (c) we chose $a=0.2$, $n_2=1$ and the localization center at $n_1=mN$ with $m=1$ --- black curve, $m=3/4$ --- magenta curve and $m=1/2$ ---cyan curve.
Notice that $I$ scales exponentially with $N$ in all cases, and the exponential rate increases with increasing slope $m$.
Finally, in panel (d) we chose $n_2=1$, $n_1=N$ and we present the quantity $I$ for two values of $a$.
Notice that the exponential rate increases with increasing $a$.

%#####################################################################################################################################

\section{Random lattice with localized eigenstates away from the edges}

\begin{figure}[h!]
\begin{center}
\includegraphics[width=0.85\columnwidth]{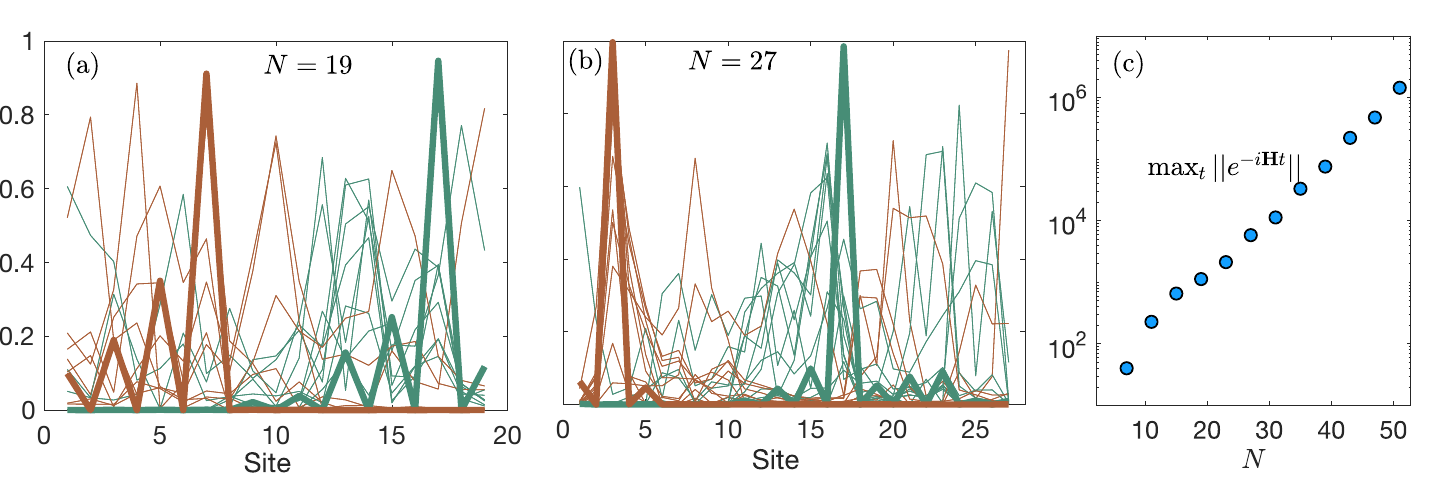}
\caption{
(a) Right (green lines) and left (brown lines) eigenstates of the random model (the bold lines indicate the zero-energy  eigenstates) for lattice size $N=19$.
We chose a realization such that the pair of zero-energy right/left eigenstates is localized at sites that satisfy $d_{n_1,n_2}=(N+1)/2$ where $d_{n_1,n_2}$ is the distance between the two localization centers.
(b) Same as (a) for $N=27$.
(c) Maximum value of the propagator norm of this random model, as a function of lattice size $N$.
}
\label{fig3_1}
\end{center}
\end{figure}

For the analysis performed in the main text, we have selected a particular realization for the random model, for each $N$, where the zero-energy right (left) eigenstate is localized at the right (left) lattice end.
That is, the center of localization of the right (left) zero-energy eigenstate is the lattice site $n_1=N$ ($n_2=1$), and therefore the distance between the two localization centers is $d_{n_1,n_2}=N-1$.
We could however, have chosen this pair of eigenstates {not to be localized at the two lattice ends} and still achieve exponential sensitivity. Such an example is illustrated in Fig.~\ref{fig3_1}.
In particular, we choose here the pair of zero-energy right/left eigenstates to be localized at sites that satisfy $d_{n_1,n_2}=(N+1)/2$ for $N=7,11,15,...$.
In Figs.~\ref{fig3_1}(a) and (b) we present the corresponding right (green lines) and left (brown lines) eigenstates of a lattice with $N=19$ and with $N=27$ respectively, under the latter constraint.
The bold lines indicate the zero-energy right/left eigenstates.
Figure~\ref{fig3_1}(c) shows the quantity $\max_t ||e^{-i\mathbf{H}t}||$ as a function of lattice size $N$, which scales exponentially with $N$.

%#####################################################################################################################################

\section{Structured pseudospectra}
\begin{figure}[h!]
\begin{center}
\includegraphics[width=1\columnwidth]{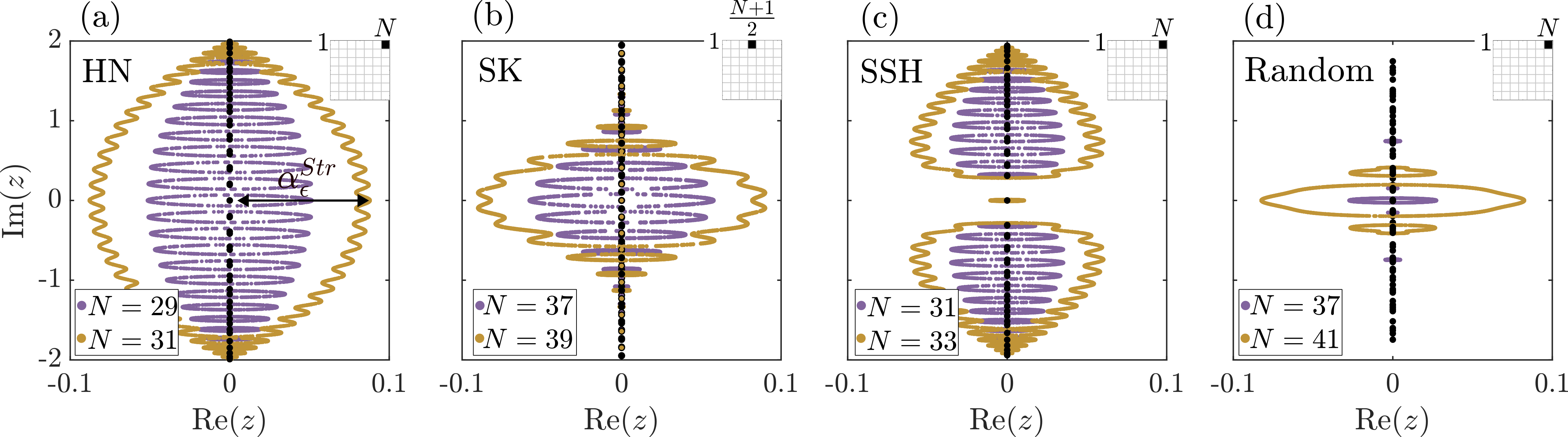}
\caption
{
Each panel corresponds to one of the four considered models.
(a)-(d) Structured pseudospectra of the matrix $\mathbf{A}=-i\mathbf{H}$ for {$10^{-5}$} and for two lattice sizes $N$.
The black dots denote the eigenvalues of $\mathbf{A}$.
The inset matrices indicate which element of the perturbed matrix $\mathbf{E}$ is nonzero.}
\label{supple_2}
\end{center}
\end{figure}

The only difference between the structured pseudospectrum \cite{Lubich2024} and the complex one is that the random matrix $\mathbf{E}$ exhibits specific structure within the former case, 
$
\sigma_{\epsilon}^{\text{Str}}(\mathbf{A})
=
\bigcup_{\mathbf{E}-\text{Structured},||\mathbf{E}|| < \epsilon}
\sigma(\mathbf{A}+\mathbf{E})
.
$
For instance, $\mathbf{E}$ may have nonzero elements exclusively on its diagonal, suggesting that only the onsite potential of the lattice is perturbed.
Below, we assume that only one element of $\mathbf{E}$ is nonzero -- a single element of the Hamiltonian is perturbed.
This single nonzero element of $\mathbf{E}$ is a complex random number of norm $\epsilon$.
Figure~\ref{supple_2} shows the eigenvalues of 300 matrices $\mathbf{A}+\mathbf{E}$ ($\mathbf{A}=-i\mathbf{H}$),
where in the HN, SSH and random models the nonzero element of $\mathbf{E}$ is the $(1,N)$,
while in the SK model the corresponding nonzero element of $\mathbf{E}$ is the $(1,\frac{N+1}{2})$.
Two lattice sizes $N$ are considered in each case and $\epsilon$ is set at {$10^{-5}$}.

Notice that for small lattice sizes (in particular for $N\epsilon \ll 1$)
the structured pseudospectrum takes the form of $N$ closed curves,
each curve surrounding an eigenvalue.
As the lattice size increases, these closed curves expand in the complex plane, eventually merging into a single curve.
We can measure the extend of the structured pseudospectrum in a manner analogous to the complex pseudospectrum,
through the structured abscissa  $\alpha_{\epsilon}^{\text{Str}}=\max \Re [\sigma_{\epsilon}^{\text{Str}}(\mathbf{A})]$.

\section{Influence of the couplings strengths to the sensitivity}

\begin{figure}[h!]
\begin{center}
\includegraphics[width=0.7\columnwidth]{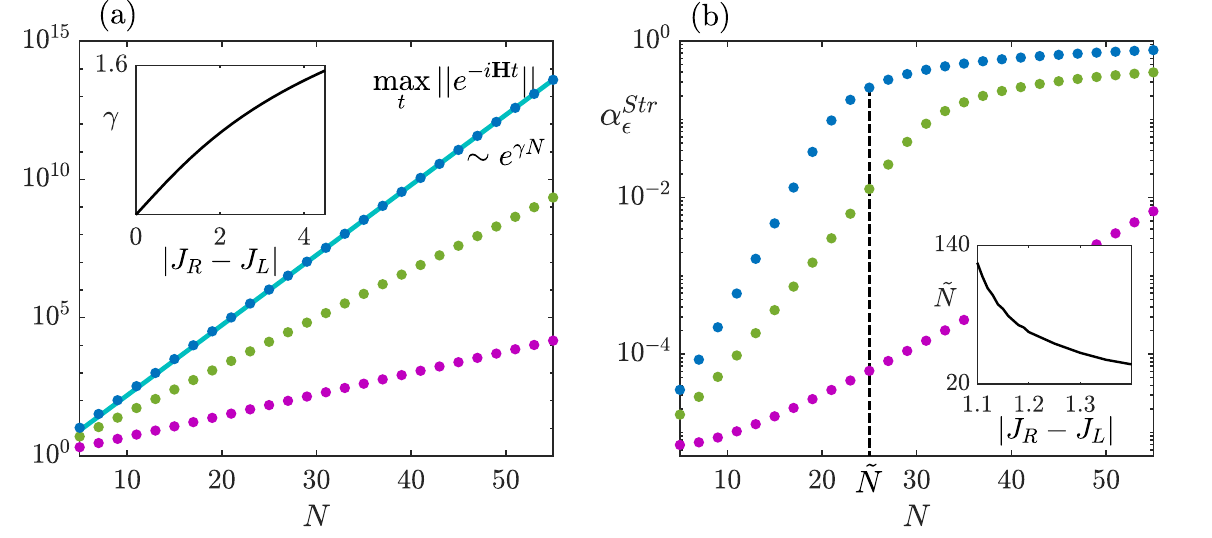}
\caption
{
(a) Maximum value of the propagator norm as a function of the lattice size $N$ for the Hatano-Nelson model and for three different couplings strengths $J_R$ and $J_L$: $J_R = 1.2$ (magenta dots), $J_R = 1.5$ (green dots), and $J_R = 1.8$ (blue dots), while $J_L = 1/J_R$ in all cases.  
The $\max_t ||e^{-i\mathbf{H}t}||$ scales as $e^{\gamma N}$.
Inset: Variation of $\gamma$ with $|J_R-J_L|$.
(b) Variation of the structured abscissa with $N$, when the $(1,N)$ element of the Hamiltonian is perturbed by $\epsilon=10^{-5}$, for the three values of $J_R,J_L$ used in panel (a).
The abscissa scales exponentially with $N$ up to some lattice size $\tilde{N}$.
Inset: Variation of $\tilde{N}$ with $|J_R-J_L|$.
}
\label{fig1_1}
\end{center}
\end{figure}

In order to illustrate the influence of the coupling strengths to the sensitivity, we present in Fig.~\ref{fig1_1}(a) the maximum value of the propagator norm $\max_t ||e^{-i\mathbf{H}t}||$ as a function of the lattice size $N$, for the Hatano-Nelson model and for three different values of hopping amplitudes $J_R$ and $J_L$ (in all cases, the condition $J_L = 1/J_R$ is imposed): $J_R = 1.2$ (magenta dots), $J_R = 1.5$ (green dots), and $J_R = 1.8$ (blue dots).
Figure~\ref{fig1_1}(a) illustrates that as the asymmetry between the right and left hopping amplitudes increases, the exponential rate also grows;
the scaling of the exponential rate $\gamma$ with the magnitude $|J_R-J_L|$ is presented in the inset of Fig.~1(a). 
Furthermore, in Fig.~\ref{fig1_1}(b) we present the structured abscissa $\alpha^{Str}$ as a function of $N$, for the three values of $J_R$ and $J_L$ used in panel (a). 
As is shown, the value of $\tilde{N}$ for which transition to the non-exponential $N$-region occurs, drops with increasing $|J_R-J_L|$ (shown in the inset of panel (b) is the dependence of $\tilde{N}$ with $|J_R-J_L|$).

%#####################################################################################################################################

\pagebreak

\section{Variation of the pseudospectral abscissa with the perturbation parameter}

\begin{figure}[h!]
\begin{center}
\includegraphics[width=1\columnwidth]{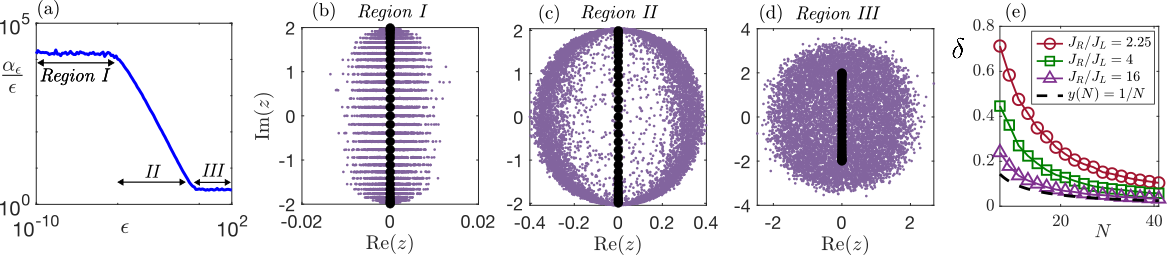}
\caption
{
(a) Variation of the abscissa with the perturbation parameter $\epsilon$, for the Hatano-Nelson model ($N=31$, $J_R=1.5$ and $J_L=1/J_R$).
In Regions I and III the abscissa scales lineraly with $\epsilon$ and in Region II it scales as a power law, $\alpha_{\epsilon} \sim \epsilon^{\delta}$. 
(b)-(d) Corresponding pseudospectrum for (b) $\epsilon=10^{-6}$, $\epsilon=10^{-3}$ and (d) $\epsilon=1$.
(e) Dependence of the exponent $\delta$ with $N$, for three values  $J_R/J_L$. Also shown is the function $y(N)=1/N$ for comparison of the sensitivity with that of an exceptional point of order $N$.  
}
\label{fig9_2}
\end{center}
\end{figure}

In Fig.~\ref{fig9_2}(a) we present the variation of the abscissa with the perturbation parameter $\epsilon$, for the Hatano-Nelson model,
considering values of $\epsilon$ ranging from low ($10^{-10}$) to high ($10^{2}$).
Notice that both x and y axis are in logarithmic scale and that the y-axis represents the abscissa over $\epsilon$:
\textit{Region I}: for $\epsilon$ approximately below $10^{-5}$, the abscissa scales linearly with $\epsilon$.
\textit{Region II}: for $10^{-5} \leq \epsilon \leq 10^0$, the abscissa scales as a power law, $\alpha_{\epsilon} \sim \epsilon^{\delta}$. 
\textit{Region III}: for $\epsilon$ beyond $10^{0}$, the abscissa scales again linearly with $\epsilon$.
Shown in Figs.~\ref{fig9_2}(b)-(d) are the spectra (black dots) and pseudospectra (purple dots) for three choices of $\epsilon$, each choice belonging in a different \textit{Region}.
Notice that in Region I the pseudospectrum consists of closed disks around each eigenvalue.
Passing to Region II, the pseudospectrum combines into a single cloud.
The variation of the exponent $\delta$ with $N$ is shown in Fig.~\ref{fig9_2}(e), for three values of asymmetry $|J_R-J_L|$.
Notice that with increasing asymmetry, the exponent $\delta(N)$ approaches the curve $1/N$.
Thus for large asymmetry, the lattice displays the same sensitivity with that of an exceptional point of order $N$.

%#####################################################################################################################################

\section{Exponential sensitivity of a $\mathcal{PT}$-symmetric lattice in the broken phase}

\begin{figure}[h!]
\begin{center}
\includegraphics[width=1\columnwidth]{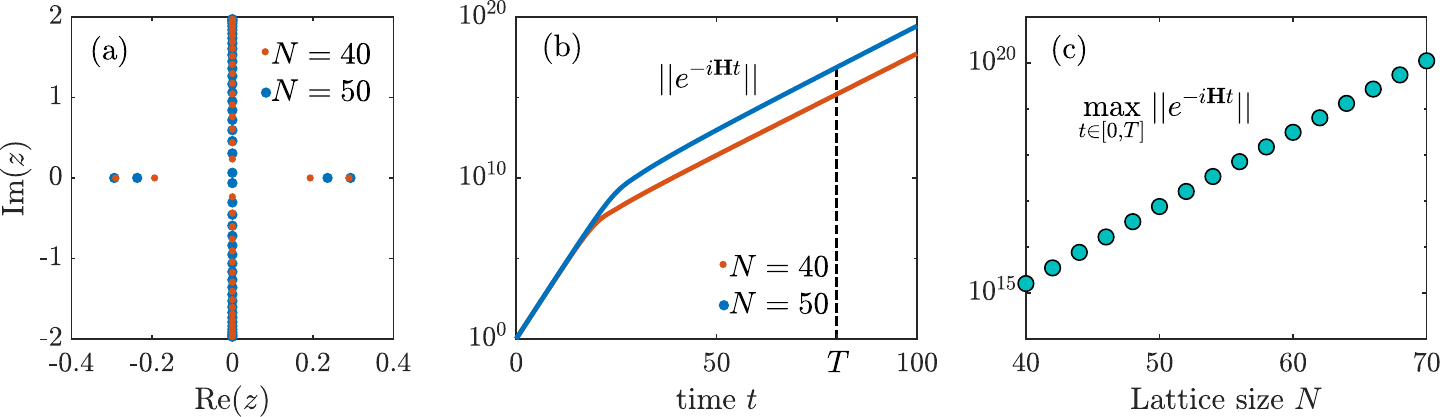}
\caption
{
(a) Eigenvalues of the matrix given in Eq.~\eqref{eq4.3} for $g=0.3$ and for two different sizes, $N=40$ and $N=50$.
(b) Time evolution of the propagator norm for the two matrices used in panel (a).
(c) Maximum value of the propagator norm as a function of size $N$.
}
\label{figsupple7n}
\end{center}
\end{figure}

We show here that a $\mathcal{PT}$-symmetric lattice with balanced gain and loss, displays exponential sensitivity with $N$, even in the $\mathcal{PT}$-broken phase.
To illustrate this, we consider the Hatano-Nelson model with alternating gain and loss at each lattice site. 
The Hamiltonian matrix describing this model is 
\begin{equation}
\mathbf{H} =
-
\begin{pmatrix}
    ig & J_L & 0  & \cdots & 0 & 0 \\
    {J}_R & -ig & J_L & \cdots & 0 & 0 \\
     & \ddots &  & & \ddots &  \\
    0 & 0 & 0  & \cdots & ig & J_L \\
    0 & 0 & 0  & \cdots & {J}_R & -ig \\
\end{pmatrix}
\label{eq4.3}
\end{equation}
where $g<0$ ($g>0$) corresponds to gain (dissipation) in the lattice site. 
Figure~\ref{figsupple7n}(a) displays the spectrum of the matrix $\mathbf{A}=-i\mathbf{H}$, for two lattice sizes $N$ and for $g=0.3$; the spectra are partially imaginary.
Shown in Fig. ~\ref{figsupple7n}(a) is the time-evolution of the propagator norm, i.e., the quantity $||e^{-i\mathbf{H}t}||$. 
Notice that $||e^{-i\mathbf{H}t}||$ increases exponentially with time, because the eigenvalues of the Hamiltonian have entered in the gainy region of the complex plane. 
Finally, in Fig.~\ref{figsupple7n}(c) we see the maximum value of the propagator norm up to some final time $T$, namely the quantity $\displaystyle \max_{t \in [0,T]}||e^{-i\mathbf{H}t}||$, as a function of $N$. Notice that it scales exponentially with $N$.

%#####################################################################################################################################

%#####################################################################################################################################

%#####################################################################################################################################


\begin{thebibliography}{99}
\bibitem{comment_norm} The matrix 2-norm is defined as $\displaystyle ||\mathbf{A}||=\max_{\boldsymbol{\psi}_0} \frac{||\mathbf{A}\boldsymbol{\psi}_0||}{||\boldsymbol{\psi}_0||}$ where $||\boldsymbol{\psi}_0||$ is the Euclidean norm of the vector $\boldsymbol{\psi}_0$.
\end{thebibliography}
\end{document}